\documentclass[a4paper]{jpconf}
\usepackage{graphicx}

\usepackage{graphicx}
\usepackage{caption}
\usepackage{amsmath}
\usepackage{cite}
\usepackage{blkarray}
\usepackage{kbordermatrix} 
\renewcommand{\kbldelim}{(} 
\renewcommand{\kbrdelim}{)}

\usepackage{xcolor}

\begin{document}
\title{A three tile 6--fold golden--mean tiling}

\author{Sam Coates$^1$, Toranosuke Matsubara$^2$, Akihisa Koga$^2$}

\address{$^1$ Department of Materials Science and Technology, Tokyo University of Science, Katsushika City, Tokyo 125--8585, Japan,\\ $^2$ Department of Physics, Tokyo Institute of Technology, Meguro, Tokyo 152--8551, Japan}

\ead{samcoates.1992@gmail.com}

\begin{abstract}
 We present a multi--edge--length aperiodic tiling which exhibits 6--fold rotational symmetry. The edge lengths of the tiling are proportional to 1:$\tau$, where $\tau$ is the golden mean $\frac{1+\sqrt{5}}{2}$. We show how the tiling can be generated using simple substitution rules for its three constituent tiles, which we then use to demonstrate the bipartite nature of the tiling vertices. As such, we show that there is a relatively large sublattice imbalance of $1/[2\tau^2]$. Similarly, we define allowed vertex configurations before analysing the tiling structure in 4--dimensional hyperspace. 
\end{abstract}

\section{Introduction}
Aperiodic tilings are key complementary tools in studying quasicrystals. Acting as model quasicrystalline structural systems in 2-- or 3--D, it is possible to explore the theoretical physical properties of tilings and to then draw parallels in their physical counterparts \cite{Takemori2020physical}. Similarly, analysing quasicrystals at the atomic level using tiling analogues allows us to build a detailed picture of structural nuances, including, for example, local phason behaviour \cite{Edagawa2000high}.

Tilings which do not share rotational symmetries with periodic systems are by far the most popular to explore, e.g.: 5--, 8--, and 12--fold. This is likely driven by their aesthetic \cite{Grunbaum1987tilings}, the overlap with physical quasicrystalline structures (in the 5-- and 12--fold cases e.g. \cite{Bursill1985penrose,Audier1986al4mn,Guyot1991quasicrystals,Ledieu2001tiling,Forster2013Perovskite}), and the initial intrigue of the exhibition of perfect long--range order in structures with `forbidden' rotational symmetry, or, orders of symmetry not associated with periodicity. However, neither quasicrystals nor aperiodic tilings are restricted to displaying non--periodic rotational symmetries \cite{Lifshitz2003quasicrystals}, and indeed, there are examples of aperiodic tilings which share symmetries with periodic systems \cite{Sasisekharan1989non,Clark1991quasiperiodic,Lifshitz2002square,Socolar2011aperiodic,Dotera2017bronze}. Further definitions of similar tilings represents an exciting arena of research: both the exploration of the physical properties of these stand--alone tilings, and the possibility of investigating interfacial periodic--aperiodic systems which share rotational symmetries.

In recent papers we presented a multi--length--scale hexagonal (6--fold) bipartite tiling with eight tiles, where the edge lengths of the tiles are governed by the golden mean, $\tau$ = $\frac{1+\sqrt{5}}{2}$ $\approx$ 1.618... \cite{Coates2022hexagonal,Koga2022ferrimagnetically}. Here, we introduce another so--called golden--mean tiling (GMT) with 6--fold symmetry and a bipartite structure which is distinct from this previous work. In the following sections we will present and  discuss the constituent tiles and their substitution rules, allowed vertex configurations, the bipartite sublattice imbalance, and the tiling structure in higher--dimensional space. We will also briefly compare our new tiling to our previous work: for clarification, we will refer to each tiling by the number of constituent tiles i.e., 8--GMT (previous) or 3--GMT (this work). 

\section*{Substitution rules}

\begin{figure}
	\centering
	\includegraphics[width=.7\linewidth]{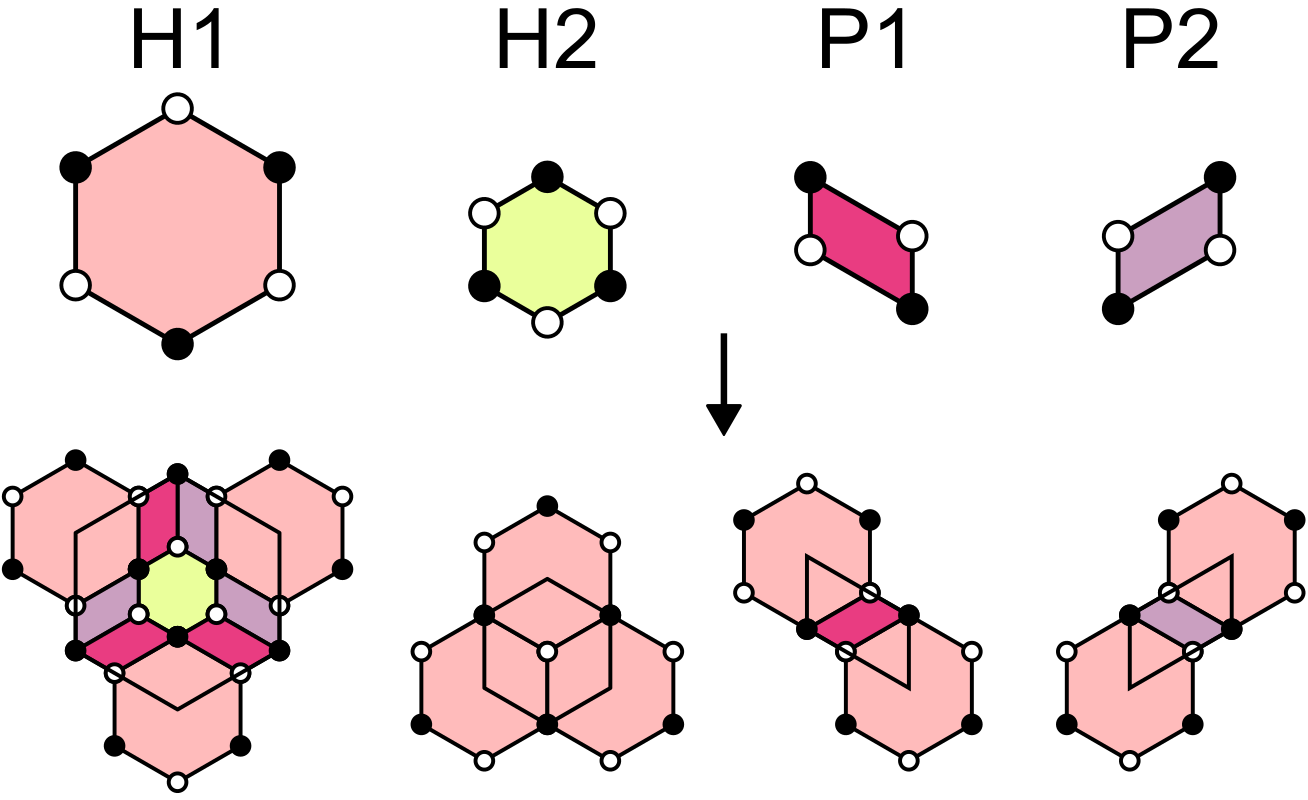}
	\caption{\textbf{Top}: 
	The four constituent tiles of the 6--fold 3--GMT, a large hexagon (H1), small hexagon (H2), and two parallelograms (P1, P2), 
	where P1 and P2 are equivalent under the mirror operation.
	 White and black circles decorate the vertices to indicate the orientational relationships of the tiles after substitution. \textbf{Bottom}: Each of the tiles after a substitution, where the outline of the original tiles is overlaid. \label{fig:deflation}}
\end{figure}

The top of Figure \ref{fig:deflation} shows the four constituent tiles of the 3--GMT, 
which are labelled with respect to their geometry: a large hexagon (H1), small hexagon (H2), and two parallelograms (P1 and P2),
where P1 and P2 are equivalent under the mirror operation.
The H1 tile has edge lengths = $\tau$, the H2 tile has edge lengths = 1, 
and the long and short sides of P1 and P2 are $\tau$ and 1 respectively, 
where the two parallelogram tiles are also found in the hexagonal 8--GMT \cite{Coates2022hexagonal}. 
Here, each of the tile vertices are decorated with either white (open) or black (closed) circles, 
to both indicate relative orientation and for later discussion on vertex configurations. The bottom of Figure \ref{fig:deflation} shows the substitution rules for each of the tiles. The outline of the original tiles are overlaid to indicate the relative relationship between tile and substitution, and the vertices are again decorated with circles. For example, we now see that the three resultant H1 tiles of the H1 tile substitution are rotated by 60$^\circ$ relative to its `parent' tile.

\begin{figure}
	\centering
	\includegraphics[width=.7\linewidth]{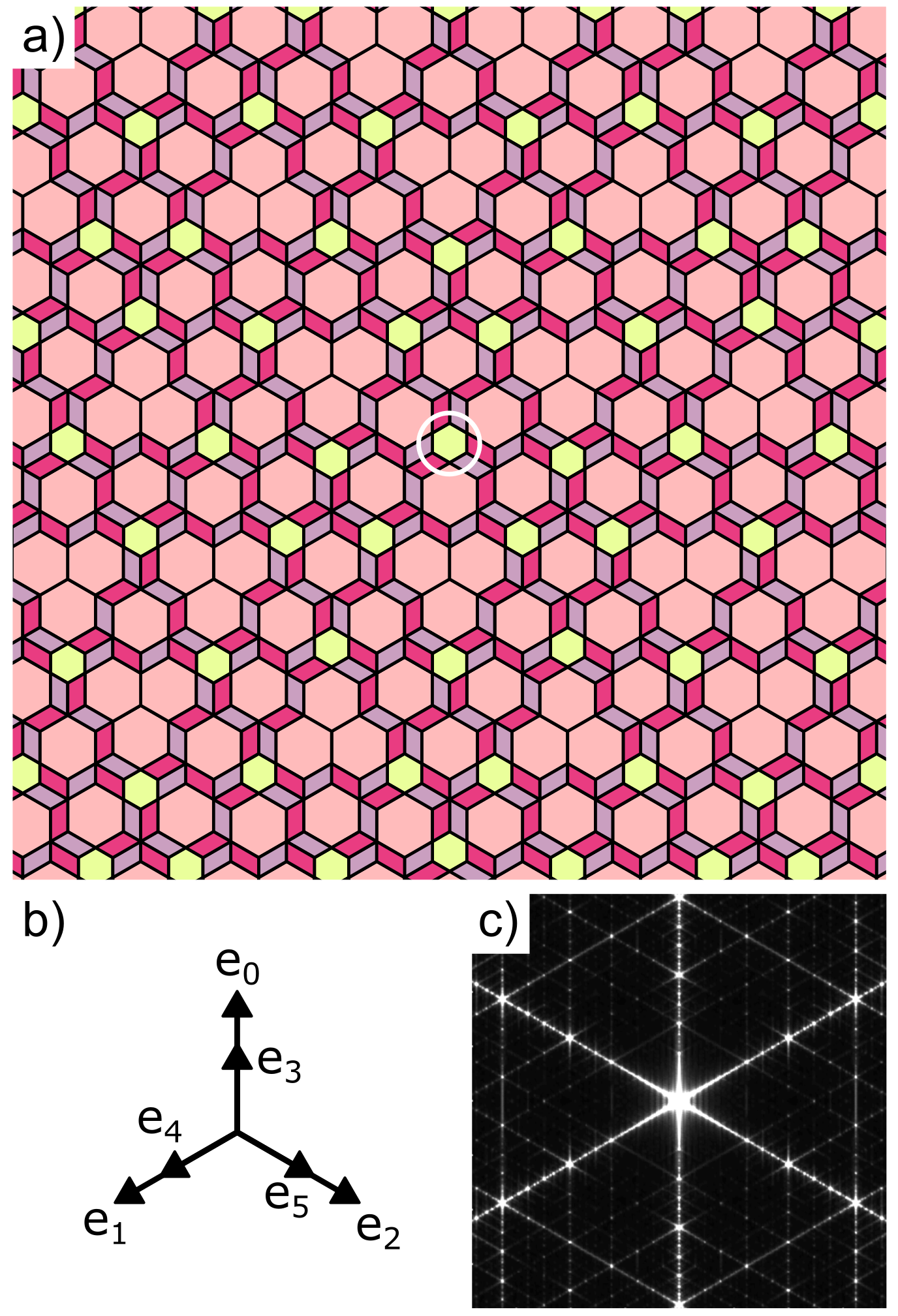}
	\caption{(a) The 6--fold 3--GMT after several generations of substitution,
          initiated from a single H1 tile.
          (b) ${\bf e}_0$, ${\bf e}_1$, $\cdots$, ${\bf e}_5$ are 
          the real space vectors. 
          (c) FFT image, which demonstrates 6--fold symmetry. \label{fig:FFT}}
\end{figure}

Figure~\ref{fig:FFT}(a) shows the 6--fold 3--GMT after a number of generations,
starting from a single H1 tile,
such that its centre of rotational symmetry is highlighted with a white circle.
To more precisely demonstrate the long--range order and the symmetry of the tiling,
we calculate its fast Fourier transform (FFT).
First, we define a set of real space  (\textbf{r}) vectors
$\text{\textbf{e}}_m=(\sin \theta, \cos \theta)$ for $m = 0,1,2$ and
$\text{\textbf{e}}_m=(\tau \sin \theta,\tau \cos \theta)$ for $m = 3,4,5$,
where $\theta$=$\frac{-2m\pi}{3}$, which are shown schematically in Fig.~\ref{fig:FFT}(b).
Then, we define reciprocal space basis vectors such that $\text{\textbf{e}}_m$$\cdot$\textbf{q}$_{m'}$ = 2$\pi\delta_{mm'}$, where $\delta_{mm'}$ is the Kronecker delta, and $\text{\textbf{e}}_m$ and \textbf{q}$_m'$ are the real and reciprocal space vectors. If each position is described by a delta function so that the density is $p(\textbf{r}) = \sum_{k=1}^{N}\delta(\textbf{\text{r}}-\textbf{\text{x}}(k))$, then the Fourier transform of the density is $p(\textbf{q}) = \sum_{k=1}^{N}\text{exp}(-i\textbf{\text{q}}\cdot\textbf{\text{x}}(k))$.
The resultant pattern is shown in Fig.~\ref{fig:FFT}(c), which
clearly demonstrates 6--fold symmetry.

Finally, we calculate the relative frequency of tiles across the infinite tiling. To do so, we introduce the substitution matrix, $M$, where column headings refer to the tile undergoing substitution, and the rows indicate how many (total) tiles of each type are contained within the substitution.
\begin{equation} 
M = 
\kbordermatrix{
	& H1 & H2 & P1 & P2 \\
	H1 & 1  & 1 & \frac{1}{3}  & \frac{1}{3}  \\
	H2 & 1  & 0 & 0  &  0  \\
	P1 &  3  &  0 & 1  & 0 \\
	P2 & 3  & 0 &  0 &  1 
}
\end{equation}
The tile frequency is found by inspecting the eigenvector of the largest eigenvalue of $M$ \cite{Lifshitz2002square,Coates2022hexagonal}, which here is $\tau^2$. The eigenvector corresponding to $\tau^2$ is $u_1$, which is found to be proportional to:
\begin{equation} 
u_1 = 
\kbordermatrix{
	\\
	H1 & \tau^2  \\
	H2 &  1  \\
	P1 &  3 \tau \\
	P2 &  3 \tau
}
\end{equation}

\noindent such that H2 is the least frequent tile, H1 is seen $\tau^2$ more often, and the two P tiles are the most frequent. The tile frequency behaviour is similar for the 3--GMT and the 8--GMT; the P tiles are seen more frequently, and there more `large' (H1) tiles than `small' (H2) by some factor of $\tau$ \cite{Coates2022hexagonal}.

\section*{Vertex matching rules}

The 6--fold 3--GMT has seven distinct vertex configurations. Figure \ref{fig:Vertices} shows each of the allowed vertices, where the top row [A, B, C] are centred on white circles, and the bottom row [D, E, F, G] are centred on black circles. 
As the prototiles are polygons with even edges, the tiling is bipartite, 
where the points of the two sublattices are centred on either white or black circles. We can calculate the relative sublattice frequency or fraction of vertex types across the infinite tiling by considering the effect of substitution on each of the allowed vertices. For example, substituting the constituent tiles of vertex A creates the following (relative) number of vertices: A $\rightarrow$ $\frac{3}{2}$B + $\frac{3}{2}$C + F + $\frac{3}{2}$G. Considering this, we generate the following matrix which describes the relative number of vertices created between generations \textit{n} and $n+1$, for each of the allowed vertex configurations:

\begin{equation} 
\centering
\kbordermatrix{
	\\
	&\text{A}_{n+1}\\
	&\text{B}_{n+1}\\
	&\text{C}_{n+1}\\
	&\text{D}_{n+1}\\
	&\text{E}_{n+1}\\
	&\text{F}_{n+1}\\
	&\text{G}_{n+1}
} = 
\kbordermatrix{
	&   &   &   & & & &  \\
	& 0  & 0 & \frac{1}{6}  & 0 &0  &0 & \frac{1}{6} \\
	& \frac{3}{2}   & 1 & \frac{1}{2}  & \frac{3}{2} &\frac{3}{2} & \frac{3}{2}&1\\
	& \frac{3}{2}   & \frac{1}{2}  & 0  & 1 & \frac{1}{2} &0 & \frac{1}{2} \\
	& 0  & 0 &  1 &  0 & 0&0 &0 \\
	& 0  & 1 &  0 &  0 & 0&0 &0\\
	& 1  & 0 &  0 &  0 & 0&0 &0\\
	&  \frac{3}{2}  & \frac{1}{2}  & 0  &  1&\frac{1}{2}  &0 &\frac{1}{2} 
}
\kbordermatrix{
	\\
	&\text{A}_n\\
	&\text{B}_n\\
	&\text{C}_n\\
	&\text{D}_n\\
	&\text{E}_n\\
	&\text{F}_n\\
	&\text{G}_n
}
\end{equation}

\noindent Similar to our treatment of the substitution matrix \textit{M}, we can find the fractions of the vertices across the infinite tiling by inspecting the eigenvector of the largest eigenvalue of the vertex matrix. Here, the largest eigenvalue is again $\tau^2$, such that the fractions of vertices are given as:

\begin{figure}
	\centering
	\includegraphics[width=0.7\linewidth]{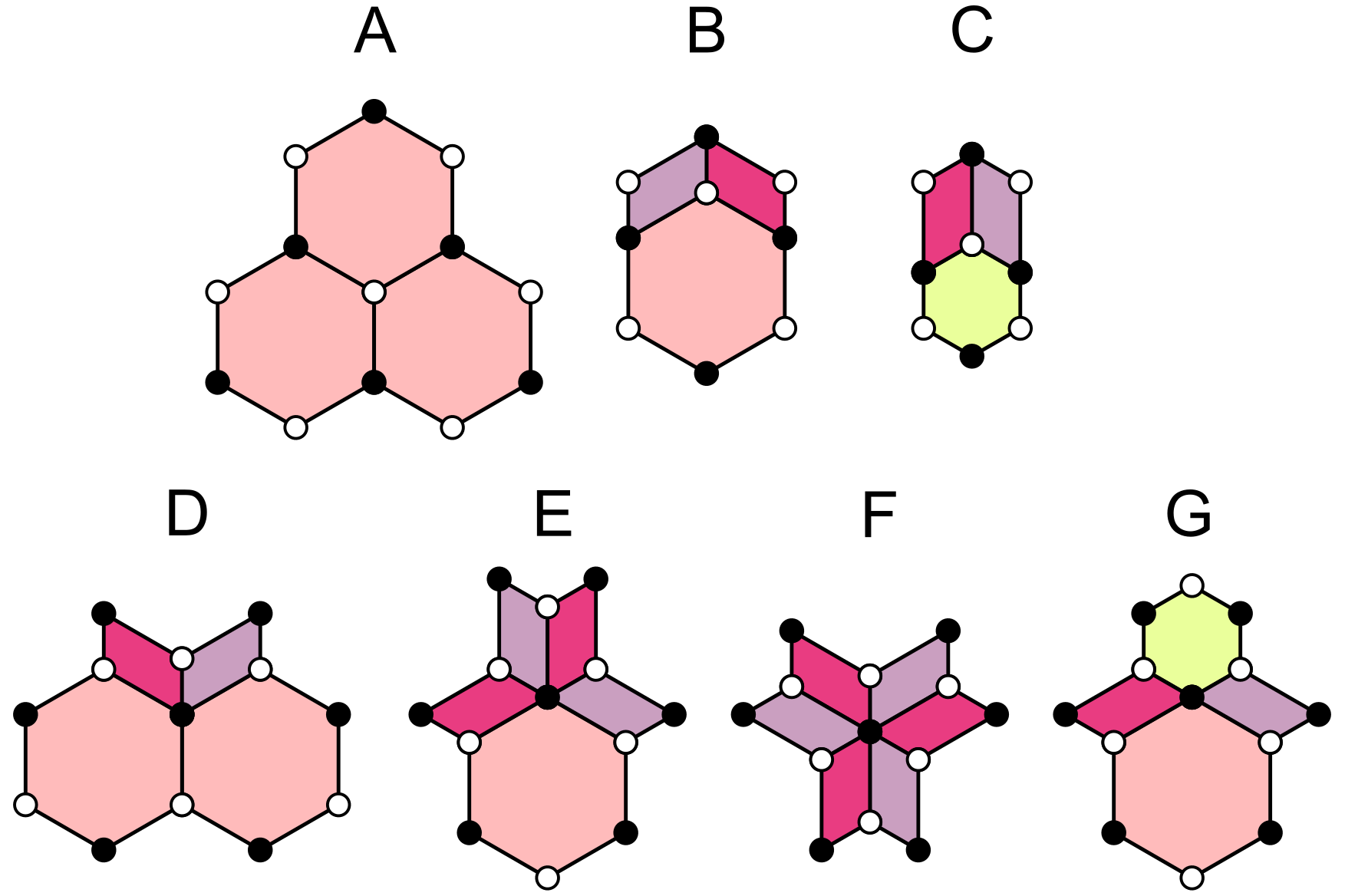}
	\caption{The seven vertex configurations of the 6--fold 3--GMT. The configurations are split into two rows, grouped by the colour of the central vertex decoration: [A, B, C], white, and [D, E, F, G], black. These groups define the bipartite sub-lattices. \label{fig:Vertices}}
\end{figure}

\begin{equation*}
\centering
\begin{aligned}
f(\text{A}) &= \frac{1}{4\tau^5} \sim 0.023\\[0.5em]
f(\text{B}) &= \frac{3 \sqrt{5}}{4\tau^3} \sim 0.395\\[0.5em]
f(\text{C}) &= \frac{3}{4\tau^3} \sim 0.177\\[0.5em]
f(\text{D}) &= \frac{3}{4\tau^5} \sim 0.068\\[0.5em]
f(\text{E}) &= \frac{3 \sqrt{5}}{4\tau^5} \sim 0.151\\[0.5em]
f(\text{F}) &= \frac{1}{4\tau^7} \sim 0.009\\[0.5em]
f(\text{G}) &= \frac{3}{4\tau^3} \sim 0.177
\end{aligned}
\end{equation*}

\noindent where we have confirmed these fractions by numerical calculation after 17 generations ($\sim$83 million vertices). We note that there is a sublattice imbalance in the bipartite distribution, where the sum of the white sublattice [A, B, C] $\sim 0.595$, and the black sublattice [D, E, F, G]
$\sim 0.404$, such that
\begin{equation*}
|f_\text{white}-f_\text{black}| = \frac{1}{2\tau^2} \sim 0.19
\end{equation*}

\noindent which is over twice as large as the imbalance in the hexagonal 8--GMT ($\sim$ 0.088) \cite{Koga2022ferrimagnetically}, so that we may expect large magnetisation under the half--filled Hubbard model, which will be discussed in future work.

\begin{figure}
	\centering
	\includegraphics[width=0.5\linewidth]{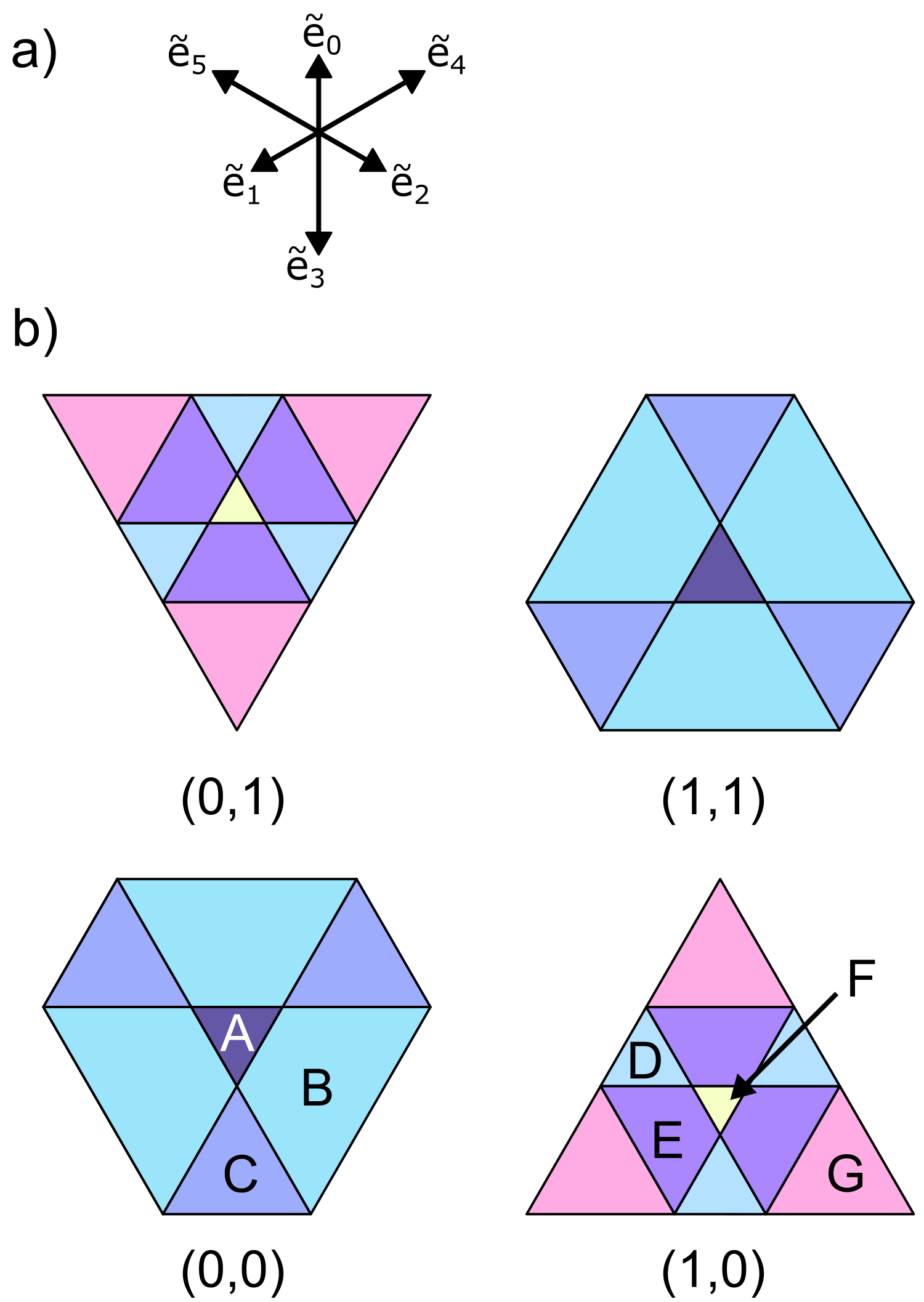}
	\caption{(a) $\tilde{\text{\textbf{r}}}$ space vectors $\tilde{\text{\textbf{e}}}_m$.
          (b) Perpendicular space $\text{\textbf{r}}^\perp$ planes for the 6--fold 3--GMT. The planes are labelled with respect to the allowed values of ($x_\perp, y_\perp$) discussed in the text.  \label{fig:Perp}}
\end{figure}

\section*{Perpendicular space}

Finally, we wish to demonstrate that the 6--fold 3--GMT point--set can be considered as a projection of a subset of points in high--dimensional space. To do so, we map the tiling to perpendicular space, where the points in perpendicular space correspond one--to--one with those in physical space. First, we note that each site on the tiling can be described by a set of six integers $\vec{n}=(n_0, n_1, ...n_5)$, so that the coordinates of the tiling in two--dimensional physical space are given by:

\begin{equation}
\text{\textbf{r}} = \sum_{m = 0}^{5} n_m\text{\textbf{e}}_m	
\end{equation}

\noindent where $\text{\textbf{e}}_m$ are the projected physical space vectors discussed above and shown in Figure \ref{fig:FFT}(b). We can then map these points onto a four--dimensional perpendicular space, which is in turn split into two two--dimensional spaces, $\tilde{\text{\textbf{r}}}$ and  $\text{\textbf{r}}^\perp$, such that $\tilde{\text{\textbf{r}}}$ points densely cover certain planes in $\text{\textbf{r}}^\perp$ \cite{Socolar1989simple,Coates2022hexagonal}. The $\tilde{\text{\textbf{r}}}$ and  $\text{\textbf{r}}^\perp$ points are given as:

\begin{equation}
\tilde{\text{\textbf{r}}} = \sum_{m = 0}^{5} n_m\tilde{\text{\textbf{e}}}_m	
\end{equation}

\begin{equation}
\text{\textbf{r}}^\perp = \sum_{m = 0}^{5} n_m\text{\textbf{e}}^\perp_m	
\end{equation}

\noindent where $\tilde{\text{\textbf{e}}}_m = \text{\textbf{e}}_{m+3}$ for \textit{m} = 0, 1, 2 and $\tilde{\text{\textbf{e}}}_m = -\text{\textbf{e}}_{m-3}$ for \textit{m} = 3, 4, 5, which are shown schematically in Figure \ref{fig:Perp}(a). $\text{\textbf{e}}^\perp_m$ = (1, 0) for \textit{m} = 0, 1, 2 and $\text{\textbf{e}}^\perp_m$ = (0, 1) for \textit{m} = 3, 4, 5. 

We find that for the sites of the 6--fold 3--GMT, $\text{\textbf{r}}^\perp$ only takes 4 values: ($x_\perp, y_\perp$) =  (0, 0), (1, 0), (0, 1) and (1, 1). The planes which are represented in these spaces by $\tilde{\text{\textbf{r}}}$ are shown in Figure \ref{fig:Perp}(b), where each plane is labelled by their $\text{\textbf{r}}^\perp$ coordinates. The planes can be considered as two sets of pairs ((0, 0), (1, 1) and (0, 1), (1, 0)), where each plane has 3--fold symmetry and is mirror symmetric to its pair. The planes can be sub--divided and coloured to represent areas which the 7 types of vertex arrangements [A, B, C, D, E, F, G] uniquely inhabit. To demonstrate, the subdivided (0, 0) and (1, 0) planes are labelled corresponding to the vertices which inhabit these areas. We note that, as expected, the fraction of the subdivided areas for each vertex type corresponds directly to the fraction of vertices in the tiling. We also note that the overall shapes of the planes are similar to four of the nine planes found in the perpendicular space construction of the 8--GMT, although they are rotated by 30$^\circ$ and found at different $\text{\textbf{r}}^\perp$ values. 

\section*{Comparison of 3--GMT and 8--GMT}

Despite any similarities, the 3--GMT and 8--GMT tilings are distinct from each other such that they are not mutually locally derivable i.e., there are no rules with which it is possible to obtain one tiling from the other \cite{Baake1991quasiperiodic}. This can be demonstrated by considering the simplest modification -- suppose that the two hexagons in the 3--GMT are decomposed into three distinct rhombuses each, such that we now have the eight tiles: two P tiles, three large rhombuses, and three small rhombuses (one can imagine an extra vertex at the centre of the two hexagonal tiles). These are now the constituent tiles of the 8--GMT, however, immediately, the vertex rules of the 8--GMT tiling are broken as three small rhombuses are not permitted to meet  \cite{Coates2022hexagonal}. Similarly, if we were to take the opposite approach of finding regions in the 8--GMT which could form large or small hexagons, we find that it is impossible to recreate, for example, vertex types A, C, or G of the 3--GMT. 

\section*{Conclusions}
Here, we presented a new 6--fold aperiodic golden--mean tiling, discussing its substitution rules, vertex configurations, bipartite lattice imbalance, and its structure in high--dimensional space. Our analysis of the tiles and vertices across the infinite tiling present intriguing possibilities, such that further studies of the confined states and magnetic properties of the 6--fold 3--GMT are currently underway.

\section*{Acknowledgements}
This work was supported by Grant-in-Aid for Scientific Research from JSPS,
KAKENHI Grants
No. JP17K05536, JP19H05821, JP21H01025, JP22K03525 (A.K.), and
No. JP19H05817 and No. JP19H05818 (S.C.).
S.C. would also like to thank Prof. Uwe Grimm for his early contributions and encouragement in the development of this tiling.

\end{document}